# TACC Technical Report TR-11-01

# Finite Element Integration on GPUs

Matthew G. Knepley
Computation Institute
University of Chicago, Chicago, IL
knepley@ci.uchicago.edu

Andy R. Terrel
Texas Advanced Computing Center
University of Texas, Austin, TX
aterrel@tacc.utexas.edu

November 1, 2018

**Abstract**

We present a novel finite element integration method for low order elements on GPUs. We achieve more than 100GF for element integration on first order discretizations of both the Laplacian and Elasticity operators.

# 1 Introduction

Graphical Processing Units (GPUs) present a promising platform for scientific simulation, offering high performance with excellent power and cost efficiency. However, despite advances in programmability for these devices [17], few numerical libraries have made use of them. The challenge of rewriting a CPU code to make use of a GPU's architectural differences is a major barrier, which can lead to slower code. As a result, high level simulation algorithms, finite elements methods (FEM) in particular, are still not widely available.

In this paper, we summarize our experience with porting general FEM integration codes from the popular FEniCS project [14] to a GPU. By adjusting the code generation tools available from FEniCS, a user is able to reuse their high level weak form definition in both a CPU or GPU code. Using our decomposition of global and local portions of the FEM integration routines, our port is able to reach up to 100 GFlops on a single machine where highly-optimized CPU codes, including hand-coded assembly routines, only reach the 12 GFlop range [5]. By creating tools that allow researchers to leverage a GPU's power throughout their code, the GPU becomes an enabler of scientific discovery rather than a limited tool for only a few codes.

We give an overview of available GPU codes for scientific computing in section 2 discussing general tactics for speeding up a code with a GPU version. For completeness, we review the tensor decomposition of FEM integration and the available form languages available from the FEniCS project in section 3. Our GPU port is described in section 4 with the numerical tests and results in section 5.

# 2 Scientific GPU Codes

Several community packages are available for basic linear algebra, such as CUBLAS [18] for the dense case and Thrust [4], CUSP [3], and CUDASparse [19] for the sparse case. While there has been excellent work bringing high order methods to the GPU, discontinuous Galerkin [12] and spectral elements [13], very little has focused on the low-order methods which make up the majority of finite element codes. Initial work in this area comes from [15], but in this paper we focus on optimizing the integration stage. Tools for runtime code generation and optimization are detailed in [11], which we will make use of in our study.

There are many excellent descriptions of the NVIDIA GPU architecture in the literature [16, 6, 17], so we will focus on the aspects salient to our problem.

GPUs can be characterized as a collection of small vector units which run in single-instruction multiple-thread (SIMT) mode. In the GTX285 model from NVIDIA on which we run our tests, the vector length is 8 and there are 30 of these Streaming MultiProcessors (SMP), as the vector units are called. In our integration implementation, we must allow enough concurrency to feed these vector units, while minimizing thread divergence and synchronization, which have large penalties on this SIMT processor. Moreover, in all GPU architectures there is a very large latency to global memory (400-600 cycles on the GTX285), as opposed to the shared and register memory co-located with the SM which cost just a few cycles. Therefore, we also minimize traffic to global memory by loading input into shared memory and storing intermediate

## 3   FEM Integration

In [8], it is shown that for any given multilinear weak form of arity $r$, we may express the element tensor as a tensor contraction,

$$A^{i_0,\ldots,i_r} = G^{\mu_0,\ldots,\mu_g} K^{i_0,\ldots,i_r}_{\mu_0,\ldots,\mu_g}. \tag{1}$$

The tensor $K$ only depends on the form itself and the reference element $\mathcal{T}_{\text{ref}}$, whereas the $G$ tensor depends on the mesh geometry and physical coefficients. Such a decomposition provides an advantage over the standard quadrature since $K$ can be precomputed and reused by all of a GPU's SMPs. The arity $g$ of $G$ depends on the transformation needed to map the form back onto the reference element, as well as any coefficients in the form.

In order to illustrate this decomposition, we will give a small example, found in Section 2 of [8]. The negative Laplacian can be expressed in weak form as

$$\begin{align}
\langle v_i, -\Delta u \rangle &= \langle \nabla v_i, \nabla u \rangle \tag{2} \\
&= \sum_e \int_{\mathcal{T}_e} \nabla v_i(\mathbf{x}) \cdot \nabla u(\mathbf{x}) d\mathbf{x} \tag{3} \\
&= \sum_e \sum_j u_j \int_{\mathcal{T}_e} \frac{\partial v_i}{\partial x_\alpha} \frac{\partial v_j}{\partial x_\alpha} d\mathbf{x} \tag{4} \\
&= \sum_e \sum_j u_j \int_{\mathcal{T}_{\text{ref}}} \frac{\partial \xi_\mu}{\partial x_\alpha} \frac{\partial v_i}{\partial \xi_\mu} \frac{\partial \xi_\nu}{\partial x_\alpha} \frac{\partial v_j}{\partial \xi_\nu} |J| d\xi. \tag{5}
\end{align}$$

where $v_i$ is any test function. Thus, the element matrix is given by

$$A_{ij} = G^{\mu\nu} K^{ij}_{\mu\nu}, \tag{6}$$

where the analytic tensor is

$$K_{\mu\nu}^{ij} = \int_{\mathcal{T}_{\text{ref}}} \frac{\partial v_i}{\partial \xi_\mu} \frac{\partial v_j}{\partial \xi_\nu} d\xi, \tag{7}$$

and the geometric tensor is

$$G^{\mu\nu} = \frac{\partial \xi_\mu}{\partial x_\alpha} \frac{\partial \xi_\nu}{\partial x_\alpha} |J| = J_{\mu\alpha}^{-1} J_{\nu\alpha}^{-1} |J|. \tag{8}$$

We have used Roman indices to indicate summation over basis functions, and Greek indices for summation over spatial dimensions.

As a second example, we express the linear elasticity operator in the same form

$$\left\langle \nabla \mathbf{v}_i + \nabla^T \mathbf{v}_i, \nabla \mathbf{u} + \nabla^T \mathbf{u} \right\rangle \tag{9}$$

$$= \sum_e \int_{\mathcal{T}_e} \left( \nabla \mathbf{v}_i + \nabla^T \mathbf{v}_i \right) : \left( \nabla \mathbf{u} + \nabla^T \mathbf{u} \right) d\mathbf{x} \tag{10}$$

$$= \sum_e \sum_j u_j \int_{\mathcal{T}_e} \left( \frac{\partial v_{i,\beta}}{\partial x_\alpha} + \frac{\partial v_{i,\alpha}}{\partial x_\beta} \right) \left( \frac{\partial v_{j,\beta}}{\partial x_\alpha} + \frac{\partial v_{j,\alpha}}{\partial x_\beta} \right) d\mathbf{x} \tag{11}$$

$$= \sum_{e,j} u_j \int_{\mathcal{T}_{\text{ref}}} \left( \frac{\partial \xi_\mu}{\partial x_\alpha} \frac{\partial v_{i,\beta}}{\partial \xi_\mu} + \frac{\partial \xi_\mu}{\partial x_\beta} \frac{\partial v_{i,\alpha}}{\partial \xi_\mu} \right) \left( \frac{\partial \xi_\nu}{\partial x_\alpha} \frac{\partial v_{j,\beta}}{\partial \xi_\nu} + \frac{\partial \xi_\nu}{\partial x_\beta} \frac{\partial v_{j,\alpha}}{\partial \xi_\nu} \right) |J| d\xi \tag{12}$$

$$\tag{13}$$

Using symmetries of this form, FFC is able to decompose this into an analytic tensor $K$

$$K_{\mu\nu}^{ij} = \frac{1}{4} \int_{\mathcal{T}_{\text{ref}}} \frac{\partial \mathbf{v}_i[\alpha]}{\partial \xi_\mu} \frac{\partial \mathbf{v}_j[\alpha]}{\partial \xi_\nu} d\xi \tag{14}$$

where $i$ and $j$ are multiindices, running over a vector valued element, and $\alpha$ is a component of this vector. The geometric tensor is identical to that for the Laplacian,

$$G^{\mu\nu} = J_{\mu\alpha}^{-1} J_{\nu\alpha}^{-1} |J|. \tag{15}$$

### 3.1 More general forms

The examples above assumed that the transformation to the reference element was affine, so that the Jacobian matrix was constant and could be removed from the element integration. If this is not the case, we may still address it using our framework by adopting the isoparametric approach. The Jacobian will be projected into a finite element space, so that

$$\int_{\mathcal{T}_{\text{ref}}} \frac{\partial \xi_\mu}{\partial x_\alpha} \frac{\partial \phi_i}{\partial \xi_\mu} \frac{\partial \xi_\nu}{\partial x_\alpha} \frac{\partial \phi_j}{\partial \xi_\nu} |J| d\xi \qquad (16)$$

$$= |J| \int_{\mathcal{T}_{\text{ref}}} \phi_k J^{-1}_{\mu\alpha,k} \frac{\partial \phi_i}{\partial \xi_\mu} \phi_l J^{-1}_{\nu\alpha,l} \frac{\partial \phi_j}{\partial \xi_\nu} d\xi \qquad (17)$$

$$= J^{-1}_{\mu\alpha,k} J^{-1}_{\nu\alpha,l} |J| \int_{\mathcal{T}_{\text{ref}}} \phi_k \frac{\partial \phi_i}{\partial \xi_\mu} \phi_l \frac{\partial \phi_j}{\partial \xi_\nu} d\xi \qquad (18)$$

$$= G^{\mu\nu}_{kl} K^{ijkl}_{\mu\nu}. \qquad (19)$$

Notice that the new coefficients, $kl$, in $G$ again depend on the particular element being integrated.

Our formalism can accomodate any multilinear operator. As a further illustration, we present the Laplace equation incorporating an inhomogeneous coefficient $w$,

$$\int_{\mathcal{T}_e} \nabla \phi_i(\mathbf{x}) \cdot w(\mathbf{x}) \nabla u(\mathbf{x}) d\mathbf{x} \qquad (20)$$

$$= \sum_{jk} u_j w_k \int_{\mathcal{T}_e} \frac{\partial \phi_i}{\partial x_\alpha} \phi_k \frac{\partial \phi_j}{\partial x_\alpha} d\mathbf{x} \qquad (21)$$

$$= \sum_{jk} u_j w_k \int_{\mathcal{T}_{\text{ref}}} \frac{\partial \xi_\mu}{\partial x_\alpha} \frac{\partial \phi_i}{\partial \xi_\mu} \frac{\partial \xi_\nu}{\partial x_\alpha} \frac{\partial \phi_j}{\partial \xi_\nu} |J| d\xi \qquad (22)$$

$$= \sum_{jk} u_j w_k G^{\mu\nu} K^{ijk}_{\mu\nu}. \qquad (23)$$

The full algebra for weak forms is detailed in [9].

Notice that the analytic $K$ tensor is an integral over products of basis functions and basis function derivatives (any member of the jet space). This means that $K$ may be calculated *a priori*, independent of the mesh or form coefficients. We will use this property to design an efficient integration method on massively parallel hardware.

### 3.2 Form Languages

Using the Unified Form Language (UFL) [1] from the FEniCS project, our system accommodates generic weak forms. We use the FEniCS Form Compiler (FFC) [9], which is implemented in Python, to process input forms and extract parts of the intermediate representation (IR) for use in GPU kernels. We illustrate this process below using linear elasticity as an example. We begin with a standard, primitive variable formulation,

$$\int_\Omega \frac{1}{4} \left( \nabla \mathbf{v} + \nabla^t \mathbf{v} \right) \cdot \left( \nabla \mathbf{u} + \nabla^t \mathbf{u} \right) d\mathbf{x} - \mathbf{v} \cdot \mathbf{f} d\mathbf{x} = 0 \qquad (24)$$

where **v** is a test function, **u** is the solution displacement, and **f** is body force. The mathematics becomes the nearly equivalent Python

```python
from ufl import interval, triangle, tetrahedron
from ufl import VectorElement, TestFunction, TrialFunction
from ufl import Coefficient, grad, inner, dx
domains = [None, interval, triangle, tetrahedron]
element = VectorElement('Lagrange', domains[dim], 1)
v = TestFunction(element)
u = TrialFunction(element)
f = Coefficient(element)

def epsilon(u):
  Du = grad(u)
  return 0.5*(Du + Du.T)

a = inner(epsilon(v), epsilon(u))*dx
L = inner(v, f)*dx
```

using the FEniCS UFL library. The FFC library can processes this form in order to extract the $G$ and $K$ tensors needed for our integration routines,

```python
import ffc
parameters = ffc.default_parameters()
parameters['representation'] = 'tensor'
analysis   = ffc.analysis.analyze_forms([a, L], {}, parameters)
ir         = ffc.compiler.compute_ir(analysis, parameters)

K = ir[2][0]['AK'][0][0].A0.astype(numpy.float32)
G = ir[2][0]['AK'][0][1]
```

where the $K$ tensor is just a numeric array, whereas the $G$ object contains instructions for constructing the geometry tensor given the element Jacobian.

## 4  GPU Implementation

Each integration kernel invocation will operate on a set of elements, which we term a *batch*, and thus the set of elements will be divided into batches, of size `elementBatchSize`, for processing. Each element integration is accomplished by contracting the geometry tensor $G$ with each block of the analytic tensor $K$, one for each element $E_{ij}$ of the element matrix. We will assign one contraction to each thread in a thread block. In order to increase concurrency, we will allow a thread block to work on multiple elements simultaneously, with the size being `numConcurrentElements`. Thus, for a vector element with dimension `numComponents` and a basis of size `numBasisFuncs`, the thread block will have $(\texttt{numBasisFuncs}*\texttt{numComponents})^2*\texttt{numConcurrentElements}$ threads.

The interleaving of computation with reads and writes to global memory is a strategy for hiding the latency of memory access. When a thread block attempts to write the result of a tensor contraction to global memory, a second thread block, currently in its compute phase, can be scheduled while it is waiting. In our experiments, shown in Section 5, interleaving resulted in noticeably higher performance, presumably due to the increased flexibility afforded to the thread block scheduler. We also employ a *thread coarsening* [21] strategy to increase performance by increasing the work per thread. This was used to great effect by Volkov [7] in his optimization of finite difference computations.

We will construct both a CPU and GPU kernel from the same source template, using the Mako [2] templating engine. This will allow us to both check the GPU results, and compare timings easily. Moreover, a single testing setup will verify both generated kernels. A similar capability could be achieved using OpenCL, specifying a different SIMT width for CPU and GPU, and more aggressive loop restructuring. This will be the focus of future work.

### 4.1 Partitioning the Computation

The integration kernel has signature

```
void integrateJacobian(int numElements,
                       float *elemMat,
                       float *geometry,
                       float *analytic)
```

on the GPU, where `geometry` is an array of the $G$ tensors for `elementBatchSize` elements, `analytic` is the $K$ tensor, and `elemMat` is an array of the element matrix for each element. On the CPU, we have

```
void integrateJacobian(int numElements,
                       float *elemMat,
                       float *geometry,
                       float *analytic)
```

where the number of elements is passed explicitly to the CPU kernel so that it can execute a loop, whereas the GPU execution grid replaces this loop. In CUDA, we use the block decomposition of kernels to partition the elements into batches,

```
/* Indexes element batch */
const int gridIdx = blockIdx.x + blockIdx.y*gridDim.x;
```

whereas on the CPU we use a loop over batches,

```
/* Loop over element batches */
const int batches = numElements/ELEMENT_BATCH_SIZE;
for(int gridIdx = 0; gridIdx < batches; ++gridIdx) {
```

where we note that in the code itself ELEMENT_BATCH_SIZE is replaced by its numeric value.

Once a batch of elements is allocated to a thread block, we assign a thread to each contraction. In CUDA, we use the thread block decomposition to index into $K$ (KROWS = numBasisFuncs*numComponents),

```
/* This is (i,j) for test and basis functions */
const int Kidx = threadIdx.x + threadIdx.y*KROWS;
/* Unique thread ID (K block is for a single element) */
const int idx  = Kidx;
```

and on the CPU we have

```
/* Loop over test functions */
for(int i = 0; i < KROWS; ++i) {
  /* Loop over basis functions */
  for(int j = 0; j < KROWS; ++j) {
    /* This is (i,j) for test and basis functions */
    const int Kidx = i + j*KROWS;
    /* Unique thread ID (K block is for a single element) */
    const int idx  = Kidx;
```

This scheme must be modified slightly when we concurrently evaluate several elements in a single thread block. In CUDA, we use the third thread block dimension to index the simultaneous evaluations,

```
/* This is (i,j) for test and basis functions */
const int Kidx = threadIdx.x + threadIdx.y*KROWS;
/* Unique thread ID
   (Same K block is used by all concurrent elements) */
const int idx  = Kidx + threadIdx.z*KROWS*KROWS;
```

and on the CPU we introduce another loop

```
/* Loop over test functions */
for(int i = 0; i < KROWS; ++i) {
  /* Loop over basis functions */
    for(int j = 0; j < KROWS; ++j) {
```

```
    /* Loop over simultaneous evaluations */
    for(int k = 0; k < NUM_CONCURRENT_ELEMENTS; ++k) {
      /* This is (i,j) for test and basis functions */
      const int Kidx = i + j*KROWS;
      /* Unique thread ID
         (Same K block is used by all concurrent elements) */
      const int idx  = Kidx + k*KROWS*KROWS;
```

Hereafter we will assume that we have simultaneous evaluations, since the reduction to the single evaluation case is straightforward. We will refer to the set of contractions performed by a given thread as the *sequential* contractions, and contractions that happen simultaneously using different sets of threads in a thread block as *concurrent* contractions. The set of threads in a thread block which all perform contractions for the same element set will be termed a *contraction set*.

## 4.2 Marshaling Data

For each sequential contraction, all threads in the contraction set must access the set of $G$ tensors for the elements in question. Therefore, these are loaded into shared memory from the `geometry` input array using a sequence of coalesced loads followed by a remainder if necessary. We illustrate this below for the case where $G$ is $3 \times 3$, `elementBatchSize` is 5, and there are 16 threads.

```
  const int       Goffset = gridIdx*DIM*DIM*ELEMENT_BATCH_SIZE;
  __shared__ float G[DIM*DIM*ELEMENT_BATCH_SIZE];

  G[idx+0]  = geometry[Goffset+idx+0];
  G[idx+16] = geometry[Goffset+idx+16];
  if (idx < 13) G[idx+32] = geometry[Goffset+idx+32];
```

In the CPU version, we merely load G from memory on the first iteration. Each thread uses a single block of $K$ for every contraction it performs. In 2D, we have, after unrolling the loop,

```
  const int Koffset = Kidx*DIM*DIM;
  float     K[DIM*DIM];

  K[0] = analytic[Koffset+0];
  K[1] = analytic[Koffset+1];
  K[2] = analytic[Koffset+2];
  K[3] = analytic[Koffset+3];
```

This load is performed after the $G$ load, but before the call to `__syncthreads()` needed to make the $G$ data available, in order to try and cover the latency of this uncoalesced read. Finally, we allocate space to hold the element matrix entry produced by each thread,

```
const int Eoffset = gridIdx*KROW*KROW*ELEMENT_BATCH_SIZE;
float     E[ELEMENT_BATCH_SIZE/NUM_CONCURRENT_ELEMENTS];
```

however we can replace `E[]` with a single scalar if we interleave calculation with writes to global storage, as shown below.

### 4.3  Computation

When computing the contraction of a set of $G$ tensors with a block of $K$, we can choose to update global memory after the entire set of contractions has been processed, or after each contraction in turn. The `interleaveStores` flag determines which strategy we pursue in the generated code. Interleaving computation with writes to global memory may allow the latency of a write to be covered by computation from another warp in the thread block, or another thread block scheduled on the SMP.

Our generation engine allows each loop to be either generated, or unrolled to produce straight-line code. In our examples, we will only display the loop code due to its brevity, but unrolled versions are presented in the results (see Section 5).

```
const int serialBatchSize =
  ELEMENT_BATCH_SIZE/NUM_CONCURRENT_ELEMENTS;
for(int b = 0; b < serialBatchSize; ++b) {
  const int n = b*numConcurrentElements;
  contractBlock('n', dim, 'E', 'G', "Goffloc", 'K', loopUnroll)
}
```

We then write each element matrix into memory contiguously with a fully coalesced write,

```
/* Store contraction results */
const int outputSize = NUM_CONCURRENT_ELEMENTS*KROWS*KROWS;
for(int n = 0; n < serialBatchSize; ++n) {
  elemMat[Eoffset+idx+n*outputSize] = E[n];
}
```

where we note that this loop is fully unrolled in the generated code.

When interleaving stores, we do a single contraction and then immediately write the result to global memory. The latency for this write can be covered by scheduling contractions in other warps on this SM. This strategy has produced consistently better results than fully calculating the contractions before writing the resulting element matrices to global memory. We show the below, where as before the contraction is fully inlined in the generated code.

```
for(int b = 0; b < serialBatchSize; ++b) {
  const int n = b*numConcurrentElements;
  E = 0.0;
  contractBlock('n', dim, 'E', 'G', "Goffloc", 'K', loopUnroll)
  /* Store contraction result */
  elemMat[Eoffset+idx+b*outputSize] = E;
}
```

## 5    Results

We demonstrate the performance of our integration method using the common Laplacian and linear elasticity operators, as shown in Fig. 5. We achieve nearly 100GF for the Laplacian, and even a little more for the elasticity operator. Note that we achieved the highest performance using interleaved stores and having each thread block operate on two elements simultaneously. The batch sizes are somewhat different, but performance is not very sensitive to this variable, as shown in Fig. 5.

To demonstrate the benefit of interleaving stores, we examine integration of the 3D $P_1$ Laplacian. The best performance was realized for an element batch size of 128 using 2 concurrent element evaluations. In Fig. 3 we show the resutls for these choices for both fully unrolled loops and the no unrolling case. Clearly, interleaving produces better performance, even in the fully unrolled case where discrepancies appear only for large runs. The disparity between the loop unrolling cases indicates that the compiler may not be applying this transformation optimally. We have performed over 3000 test runs with various combinations of the input parameters, which are archived along with the source code, so that they may be mined in a similar fashion by other researchers.

## 6    Discussion

We note that a version of the Laplace kernel was tested in which $K$ is loaded into shared memory and all threads perform the complete contraction with a given $G$ simultaneously. However, this strategy was abandoned due to lack of efficiency, mainly arising from the lower level of concurrency available.

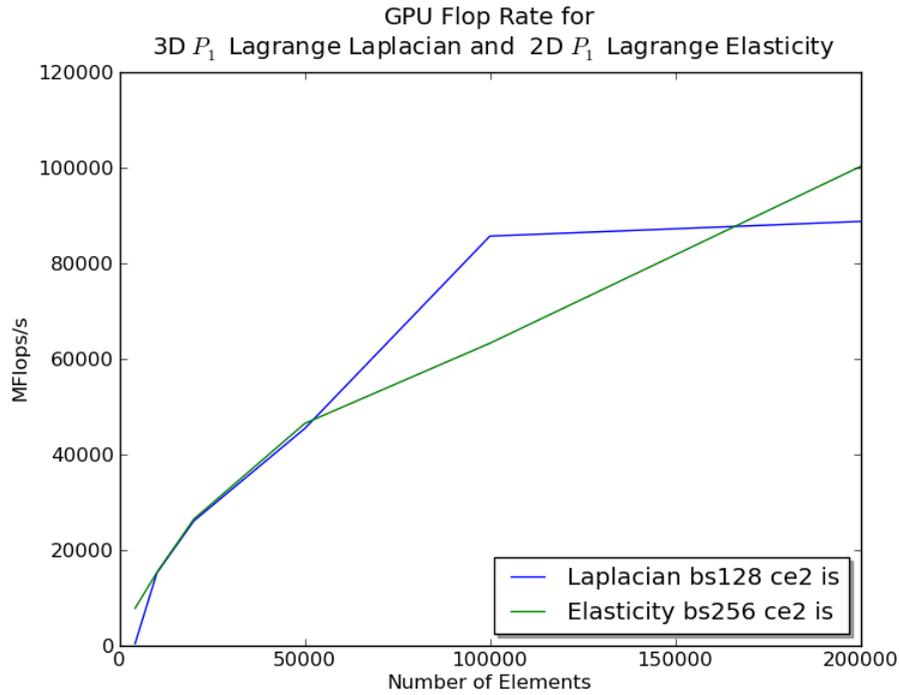

Figure 1: This graph shows the peak performance achieved for element integration of the 3D $P_1$ Laplacian and 2D $P_1$ Elasticity operators. We use *bs* to denote the element batch size, *ce* the number of concurrent element evaluations, *is* interleaved stores, and *unroll* for fully unrolled contraction loops.

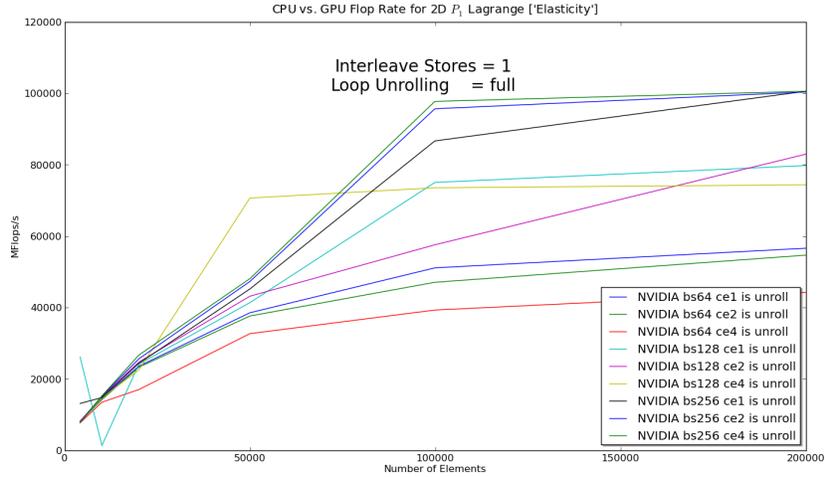

Figure 2: This graph shows the dependence of flop rate on the element batch size for the 2D $P_1$ Elasticity operator. We use *bs* to denote the element batch size, *ce* the number of concurrent element evaluations, *is* interleaved stores, and *unroll* for fully unrolled contraction loops.

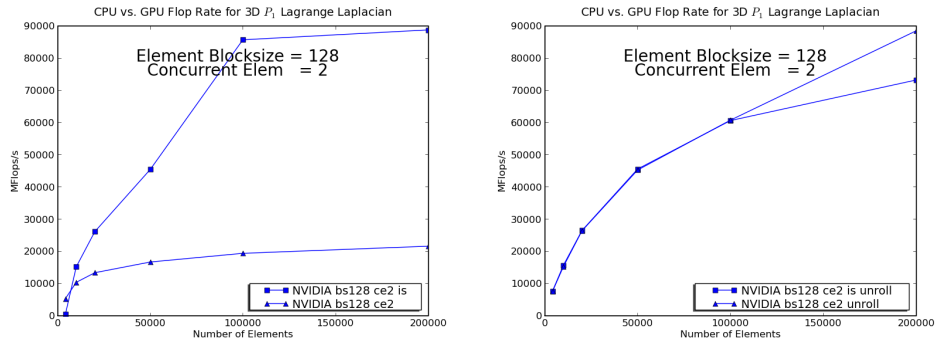

Figure 3: This graph shows the dependence of flop rate on the the interleaving of global stores for the 3D $P_1$ Laplacian operator. We use *bs* to denote the element batch size, *ce* the number of concurrent element evaluations, *is* interleaved stores, and *unroll* for fully unrolled contraction loops. The left graph shows performance with no loop unrolling, and the right for fully unrolled loops.

We will extend these initial results to more complex operators with variable coefficients, as well as systems of equations which arise in multiphysics problems. This will necessitate a more systematic approach to optimization over the algorithmic variants. We plan to use the loop slicing tool Loo.py [10] and generated, optimized quadrature rules from FFC [20] in addition to exhaustive strategies. We have an integrated build and test framework, which allows us to run all variants in a single execution and record the results in HDF5 for later processing and analysis. Moreover, when processing coupled systems, we will be able to break the weak form into blocks commensurate with different preconditioning strategies and evaluate the performance. This entire package will be integrated into both PETSc and FEniCS for easy use by projects already these frameworks.